\documentstyle[epsf,12pt]{article}
\begin{document}
\begin{titlepage}
\title{
Binding Energy of Scalar Bound State by Topologically Massive Interaction: 
Fermion and Anti-fermion System with Heavy Mass 
}
\author{
Toyoki Matsuyama \thanks{e-mail address: matsuyat@nara-edu.ac.jp} \\
Department of Physics, Nara University of Education \\
Takabatake-cho, Nara 630-8528, Japan \\
and \\
Hideko Nagahiro \\
Department of Physics, Nara Women's University \\
Nara 630-8506, Japan
}
\date{\today}
\end{titlepage}
\maketitle
\begin{abstract}
A bound state problem in a topologically massive quantum 
electrodynamics is investigated by using a non-perturbative method.  
We formulate the Bethe-Salpeter equation for scalar bound states composed of 
massive fermion and anti-fermion pair under the lowest ladder approximation.  
In a large mass expansion for the (anti-) fermion, we derive the 
Schr{\"o}dinger equation and solve it by a numerical method.  
The energy eigenvalues of bound states are evaluated for various values of 
a topological mass and also a fermion mass.  
Then we find a novel logarithmic scaling behaviour of the binding energy 
in varying the topological mass, fermion mass and also a quantum number.  
There exists a critical value of the topological mass, beyond which the bound 
states disappear.  
As the topological mass decreases, the energy eigenvalues of the bound 
states, which are negative, also decrease with a logarithmic dependence on 
the topological mass.  
A Chern-Simons term gives the bound system a repulsive effect.  
\end{abstract}

\section{Introduction}

A peculiar type of quantum field theory exists in (2+1)-dimensional 
(generally odd-dimensional) space-time.  
It is the theory that has a {\it Chern-Simons term} in the 
action~\cite{SS,DJT}.  
While the Chern-Simons term may be added to the action by hand, the term can 
be induced in an effective action by a vacuum polarization effect of a fermion 
as the anomaly~\cite{RNSI}.  
Our understanding on the theory has progressed by studying the topological 
structure due to the Chern-Simons term~\cite{TJZW} 

In addition to the theoretical interests, two important phenomena were 
discovered in {\it planar} condensed matter systems.  
One is the quantum Hall effect~\cite{QH} and the other is the high-$T_c$ 
superconductivity~\cite{HTC}.  
Both phenomena may be thought of as realizations of the macroscopic quantum 
effects in the planar systems in which electrons have a strong correlation.  
It is plausible that these phenomena may have their origins in the dimensions 
of the space-times.  
It is challenging to ask whether and how the Chern-Simons term is 
related to these phenomena discovered in the two-dimensional electron 
systems.  

These situations have activated the study of the (2+1)-dimensional quantum 
field theories.  
In fact, there appeared many approaches to understand these macroscopic 
quantum effects by using the (2+1)-dimensional quantum field theories and by 
paying special attention to the effects of the Chern-Simons term.  
For example, the (2+1)-dimensional quantum electrodynamics (QED$_3$) was used 
to explain the quantum Hall effect ~\cite{IM} and QED$_3$ with the 
Chern-Simons term (Maxwell-Chern-Simons QED$_3$) provided an anyon 
model~\cite{WSM} which was expected to give us an essential mechanism for 
the high-$T_C$ superconductivity.  
These investigations produced important results and are still in progress.

What is the physical meaning of the Chern-Simons term?
In the (2+1)-dimensional gauge theories, the behaviors in the infrared regions 
seem to be unstable at least in the perturbative treatment.  
The Chern-Simons term gives the gauge field a topological mass without 
breaking the gauge symmetry so that the term may rescue the theory from the 
infrared catastrophe.  
This is an original motivation of including the Chern-Simons term in the 
action.~\cite{JTAPRS}  

Reminding those studies, we think that one of the most important problems 
which are still unclear is what is a role of the Chern-Simons term in a 
$nonperturbative$ region.  
It is our purpose to clarify how the Chern-Simons term affects the 
nonperturbative dynamics. 
In previous works~\cite{PABCWABKW,DKK,KKKM,HMH,MNU}, a dynamical mass 
generation of a fermion has been investigated by using the Schwinger-Dyson 
equation.  
It is an important phenomenon induced by a nonperturbative effect in one-body 
system.  
In this paper, we study a bound state problem which is another important 
nonperturbative phenomenon appeared in  two-body system.  
We are interested in how the Chern-Simons term affects the bound system 
through the nonperturbative dynamics.

In Ref.~\cite{BAB}, a fermion-anti-fermion bound system in QED$_3$ was 
studied.  
In the model used there, the (anti-) fermion had four components and the gauge 
field did not have the topological mass.  
On the other hand, a binding problem of fermion-fermion system~\cite{K} was 
investigated in Ref.~\cite{GGS}.   
A two-component fermion was used and the Chern-Simons term was included 
in the action of gauge field.  
The effective fermion-fermion low-energy potential was evaluated in the 
lowest order perturbation.  
However, the result was claimed to be incorrect~\cite{H,DEKSS}.  

As mentioned above, what we are interested in is the effect of the 
Chern-Simons term in the nonperturbative dynamics.  
It is still unclear and controversial.  
We think that it is important to perform the microscopic nonperturbative 
calculation based on the first principle.  
For this purpose, the best theoretical tool is the Bethe-Salpeter equation 
which makes it possible to perform the systematic  
nonperturbative analysis of the structure of the bound system.
As a system suitable for our purpose, we use the two-component massive 
fermion and anti-fermion in QED$_3$ with the Chern-Simons term.   
This model has the topological nature and also has applications in the 
condensed matter physics.  

This paper is organized as follows: 
In Sec. 2, we explain the quantum electrodynamics with the Chern-Simons term.  
The Bethe-Salpeter equation is formulated under the lowest ladder 
approximation in Sec. 3.  
In Sec. 4, we derive the Schr\"{o}dinger equation using the large fermion mass 
expansion.  
After that, by solving the Schr\"{o}dinger equation numerically, we evaluate 
the binding energy for the bound states and we find a specific logarithmic scaling
behaviour of the binding energy in Sec. 5.  
Finally, we conclude on the results with discussions in Sec. 6.  

\section{Maxwell-Chern-Simons QED$_3$}

The Lagrangian density of QED$_3$ with the Chern-Simons term, which is called 
the Maxwell-Chern-Simons QED$_3$ or the topologically massive QED$_3$, is 
expressed as 
%%%%%%%%%%%%%%%%%%%%%%%%%%%%%%%%%%%%%%%%%%%%%%%%%%%%%%%%%%%%%%%%%%%%%%%%%%%%%
\begin{eqnarray}
{\cal L}=-\frac{1}{4}F_{\mu\nu}F^{\mu\nu}
         +\frac{\mu}{2}\varepsilon^{\mu\nu\rho}A_\mu \partial_\nu A_\rho
         -\frac{1}{2\alpha}(\partial_\mu A^\mu)^2 
        + \bar{\psi}(i \not \hspace{-0.7mm}\partial
         -e\not \hspace{-1.2mm}A-m)\psi \ , 
\label{lagrangian}
\end{eqnarray}
%%%%%%%%%%%%%%%%%%%%%%%%%%%%%%%%%%%%%%%%%%%%%%%%%%%%%%%%%%%%%%%%%%%%%%%%%%%%%
where $e$ is the coupling constant and $\alpha$ is the gauge-fixing 
parameter.  
$\psi$ is a two component fermion field with an explicit mass $m$, which 
belongs to the irreducible spinor representation in (2+1)-dimensions.  
In the dimensions, the square of the coupling constant $e^2$ has a 
dimension of mass.  
The second term on the right-hand side of Eq. (\ref{lagrangian}) is the 
so-called Chern-Simons term.  
This term has a topological meaning as a secondary characteristic 
class~\cite{TJZW,CS} and provides the topological mass $\mu$ to the gauge 
field without breaking the gauge symmetry.
The free propagator of the gauge field $iD_{\mu\nu}(p)$ derived from 
Eq. (\ref{lagrangian}) is 
%%%%%%%%%%%%%%%%%%%%%%%%%%%%%%%%%%%%%%%%%%%%%%%%%%%%%%%%%%%%%%%%%%%%%%%%%%%%%
\begin{eqnarray}
iD_{\mu\nu}(p)= -i \frac{1}{p^2-\mu^2} \left(g_{\mu\nu} 
                    - \frac{p_\mu p_\nu}{p^2} \right) 
              + \mu\frac{1}{p^2-\mu^2} \frac{1}{p^2} 
                      \varepsilon_{\mu\nu\rho}p^\rho 
                    -i\alpha\frac{p_\mu p_\nu}{p^4} \ .  
\label{gauge_pro}
\end{eqnarray}
%%%%%%%%%%%%%%%%%%%%%%%%%%%%%%%%%%%%%%%%%%%%%%%%%%%%%%%%%%%%%%%%%%%%%%%%%%%%%
We find a massive pole at $p^2=\mu^2$ in the propagator so that $\mu$ can be 
regarded as a mass of the gauge field.  

The Dirac matrices in (2+1)-dimensions are defined by 
%%%%%%%%%%%%%%%%%%%%%%%%%%%%%%%%%%%%%%%%%%%%%%%%%%%%%%%%%%%%%%%%%%%%%%%%%%%%%
\begin{eqnarray}
\gamma^0=\sigma_3, \ \gamma^1=i\sigma_1, \ \gamma^2=i\sigma_2 \ ,
\label{dirac}
\end{eqnarray}
%%%%%%%%%%%%%%%%%%%%%%%%%%%%%%%%%%%%%%%%%%%%%%%%%%%%%%%%%%%%%%%%%%%%%%%%%%%%%
with diag$(g^{\mu\nu})=(1,-1,-1)$.  
$\sigma_i$'s $(i=1,2,3)$ are the Pauli matrices.  
In this representation, there is no matrix which anti-commutes with all of 
the $\gamma$-matrices so that we cannot define the chiral transformation.  
Instead, the parity symmetry~\cite{SS,DJT} forbids the mass terms of both the 
fermion and the gauge field.  
The Lagrangian density (\ref{lagrangian}) breaks the parity symmetry in both 
the gauge and fermion sectors.  

\section{Bethe-Salpeter Equation}

The Bethe-Salpeter amplitude $\chi_B(x_a,x_b)$ is defined as 
%%%%%%%%%%%%%%%%%%%%%%%%%%%%%%%%%%%%%%%%%%%%%%%%%%%%%%%%%%%%%%%%%%%%%%%%%%%%%
\begin{eqnarray}
\chi_B^{ij}(x_a,x_b)\equiv \langle 0|
T(\psi^i(x_a)\bar{\psi}^j(x_b))|k\rangle \ ,
\label{eq:BS_amp}
\end{eqnarray}
%%%%%%%%%%%%%%%%%%%%%%%%%%%%%%%%%%%%%%%%%%%%%%%%%%%%%%%%%%%%%%%%%%%%%%%%%%%%%
where $|k\rangle$ is the fermion-anti-fermion bound state with the total 
momentum $k_\mu$ and mass $M$.
The mass shell condition is $k^2=k_0^2-{\vec k}^2=M^2$ with $0 <M < 2m$.  
The integral Bethe-Salpeter equation~\cite{BS,L} for $\chi_B(x_a,x_b)$ is given by
%%%%%%%%%%%%%%%%%%%%%%%%%%%%%%%%%%%%%%%%%%%%%%%%%%%%%%%%%%%%%%%%%%%%%%%%%%%%%
\begin{eqnarray}
\chi_B(x_a,x_b)=-e^2\int d^3z_a d^3z_b iS_F'(x_a-z_a) 
\gamma^\mu\chi_B(z_a,z_b)\gamma^\nu 
\nonumber \\
\times iS_F'(z_b-x_b) iD_{\mu\nu}'(z_a-z_b) \ ,
\label{BSeq}
\end{eqnarray}
%%%%%%%%%%%%%%%%%%%%%%%%%%%%%%%%%%%%%%%%%%%%%%%%%%%%%%%%%%%%%%%%%%%%%%%%%%%%%
where $iS_F'$ and $iD_{\mu \nu}'$ are full propagators of the fermion and 
gauge fields respectively.  

Now we limit ourselves to solve Eq. (\ref{BSeq}) in the lowest ladder 
approximation.  
Then $iS_F'$ is replaced by the free fermion propagator $iS_F$ as
%%%%%%%%%%%%%%%%%%%%%%%%%%%%%%%%%%%%%%%%%%%%%%%%%%%%%%%%%%%%%%%%%%%%%%%%%%%%%
\begin{eqnarray}
{iS_F}(x_a-x_b)=i\int\frac{d^3k}{(2\pi)^3}e^{-ik (x_a-x_b)} S_F(k) , \ \ 
{iS_F}(k)=i \frac{(\not \hspace{-0.5mm}k+m)}{k^2-m^2+i\epsilon} , 
\label{fer_pro}
\end{eqnarray}
%%%%%%%%%%%%%%%%%%%%%%%%%%%%%%%%%%%%%%%%%%%%%%%%%%%%%%%%%%%%%%%%%%%%%%%%%%%%%
and $iD_{\mu \nu}'$ by the free gauge field propagator $iD_{\mu \nu}$ as 
%%%%%%%%%%%%%%%%%%%%%%%%%%%%%%%%%%%%%%%%%%%%%%%%%%%%%%%%%%%%%%%%%%%%%%%%%%%%%
\begin{eqnarray}
iD_{\mu\nu}(x_a-x_b)=\int\frac{d^3k}{(2\pi)^3}
iD_{\mu\nu}(k)e^{-ik(x_a-x_b)}, 
\end{eqnarray}
%%%%%%%%%%%%%%%%%%%%%%%%%%%%%%%%%%%%%%%%%%%%%%%%%%%%%%%%%%%%%%%%%%%%%%%%%%%%%
where $iD_{\mu \nu}(k)$ is given by Eq. (\ref{gauge_pro}).  

The dependence of $\chi_B(x_a,x_b)$ on the coordinates of the centre of mass 
can be factored out as 
%%%%%%%%%%%%%%%%%%%%%%%%%%%%%%%%%%%%%%%%%%%%%%%%%%%%%%%%%%%%%%%%%%%%%%%%%%%%%
\begin{eqnarray}
\chi_B^{ij}(x_a,x_b)=e^{-iP X}\chi_B^{ij}(x,P),
\end{eqnarray}
%%%%%%%%%%%%%%%%%%%%%%%%%%%%%%%%%%%%%%%%%%%%%%%%%%%%%%%%%%%%%%%%%%%%%%%%%%%%%
where $X=(x_a+x_b)/2$ is the coordinates of the centre of mass and $x=x_a-x_b$ 
the relative coordinates.  
$P$ is the momentum of the centre of mass.  
By the Fourier transformation of the relative coordinates $x$, we 
obtain  
%%%%%%%%%%%%%%%%%%%%%%%%%%%%%%%%%%%%%%%%%%%%%%%%%%%%%%%%%%%%%%%%%%%%%%%%%%%%%
\begin{eqnarray}
\chi_B^{ij}(x_a,x_b)= e^{-iP X}
\int\frac{d^3p}{(2\pi)^3}e^{-ip x}\chi_B^{ij}(p,P),
\end{eqnarray}
%%%%%%%%%%%%%%%%%%%%%%%%%%%%%%%%%%%%%%%%%%%%%%%%%%%%%%%%%%%%%%%%%%%%%%%%%%%%%
where $p$ is the relative momentum.  

We may rewrite the Bethe-Salpeter equation by using an {\it amputated} Bethe-Salpeter 
amplitude $\Gamma$ which is defined as
%%%%%%%%%%%%%%%%%%%%%%%%%%%%%%%%%%%%%%%%%%%%%%%%%%%%%%%%%%%%%%%%%%%%%%%%%%%%%
\begin{eqnarray}
\Gamma(q,P)\equiv -\left[ iS_F\left(\frac{\not\hspace{-0.7mm} P}{2} 
+\not\hspace{-0.7mm} p \right) \right]^{-1}
\chi_B(p,P) \left[ iS_F \left(-\frac{\not\hspace{-0.7mm} P}{2} 
+\not\hspace{-0.7mm} p \right) \right]^{-1} .
\label{eq:Gamma}
\end{eqnarray}
%%%%%%%%%%%%%%%%%%%%%%%%%%%%%%%%%%%%%%%%%%%%%%%%%%%%%%%%%%%%%%%%%%%%%%%%%%%%%

From Eqs. (\ref{BSeq}) $\sim$ (\ref{eq:Gamma}),  we obtain 
%%%%%%%%%%%%%%%%%%%%%%%%%%%%%%%%%%%%%%%%%%%%%%%%%%%%%%%%%%%%%%%%%%%%%%%%%%%%%
\begin{eqnarray}
\Gamma(q,P)=-e^2\int\frac{d^3p}{(2\pi)^3} iD_{\mu\nu}(q-p) \gamma^\mu 
iS_F(\frac{\not\hspace{-0.5mm}P}{2}+\not\hspace{-0.5mm}p)
\Gamma(p,P) 
iS_F(-\frac{\not\hspace{-0.5mm}P}{2}+\not\hspace{-0.5mm}p)  
\gamma^\nu \ .  
\label{BS_Gamma}
\end{eqnarray}
%%%%%%%%%%%%%%%%%%%%%%%%%%%%%%%%%%%%%%%%%%%%%%%%%%%%%%%%%%%%%%%%%%%%%%%%%%%%%
Now the problem is to find the amputated Bethe-Salpeter amplitude $\Gamma$ by 
solving Eq. (\ref{BS_Gamma}).  
Because of the Lorentz invariance, the amplitude which is a bi-spinor, can be 
decomposed as  
%%%%%%%%%%%%%%%%%%%%%%%%%%%%%%%%%%%%%%%%%%%%%%%%%%%%%%%%%%%%%%%%%%%%%%%%%%%%%
\begin{eqnarray}
\Gamma(q,P)\equiv M(q,P)+N(q,P)\not\hspace{-0.5mm}q+F(q,P)\not\hspace{-1mm}P 
+G(q,P) \varepsilon^{\mu\nu\rho} q_\mu P_\nu \gamma_\rho,
\label{Gamma}
\end{eqnarray}
%%%%%%%%%%%%%%%%%%%%%%%%%%%%%%%%%%%%%%%%%%%%%%%%%%%%%%%%%%%%%%%%%%%%%%%%%%%%%
where $M,N,F$ and $G$ are the functions of the Lorentz scalar variables 
$q=\sqrt{q_\mu q^\mu}$, $P=\sqrt{P_\mu P^\mu}$ and $P q=P_\mu q^\mu$.  
The equations for the four unknown functions $M,N,F$ and $G$ can be obtained 
by taking the trace of the function $\Gamma$ as 
%%%%%%%%%%%%%%%%%%%%%%%%%%%%%%%%%%%%%%%%%%%%%%%%%%%%%%%%%%%%%%%%%%%%%%%%%%%%%
\begin{eqnarray}
M(q,P)&=&\frac{1}{2}{\rm tr}\left[\Gamma(q,P)\right] \ , 
\label{M1} \\
N(q,P)&=&\frac{1}{2}\frac{P^2q^\sigma-(P\cdot q)P^\sigma} 
{P^2q^2-(P\cdot q)^2}{\rm tr}\left[\gamma_\sigma\Gamma(q,P)\right] \ , 
\label{N} \\
F(q,P)&=&\frac{1}{2}\frac{q^2P^\sigma-(P\cdot q)q^\sigma} 
{P^2q^2-(P\cdot q)^2}{\rm tr}\left[\gamma_\sigma \Gamma(q,P)\right] \ , 
\label{F} \\
G(q,P)&=&\frac{1}{2}\frac{\varepsilon^{\alpha\beta\sigma} 
q_\alpha P_\beta}{P^2q^2-(P\cdot q)^2}{\rm tr} 
\left[\gamma_\sigma\Gamma(q,P)\right] \ .  
\label{G1}
\end{eqnarray}
%%%%%%%%%%%%%%%%%%%%%%%%%%%%%%%%%%%%%%%%%%%%%%%%%%%%%%%%%%%%%%%%%%%%%%%%%%%%%

In the following section, we rewrite the obtained four coupled integral 
equations into the non-relativistic form using the large fermion mass 
expansion.  
This leads to the Schr\"{o}dinger equation which will be used actually to 
study the bound states.  
The non-relativistic treatment reduces the complexity of the equations as we 
will see in the next section.  

\section{Schr\"odinger Equation} 

We consider the bound state in the rest frame where the momentum of the 
centre of mass is 
%%%%%%%%%%%%%%%%%%%%%%%%%%%%%%%%%%%%%%%%%%%%%%%%%%%%%%%%%%%%%%%%%%%%%%%%%%%%%
\begin{eqnarray}
P_\mu=(2m+E^B,0,0) .
\label{mom}
\end{eqnarray}
%%%%%%%%%%%%%%%%%%%%%%%%%%%%%%%%%%%%%%%%%%%%%%%%%%%%%%%%%%%%%%%%%%%%%%%%%%%%%
$E^B$ indicates the binding energy and $m$ the (anti-) fermion mass.  
If the system makes bound states, $E^B$ should be negative because of a 
stability.  
We replace all $P_\mu$'s in Eq. (\ref{BS_Gamma}) by Eq. (\ref{mom}).  
Then the fermion propagator in the right-hand side of Eq. (\ref{BS_Gamma}) can 
be approximated as 
%%%%%%%%%%%%%%%%%%%%%%%%%%%%%%%%%%%%%%%%%%%%%%%%%%%%%%%%%%%%%%%%%%%%%%%%%%%%%
\begin{eqnarray}
\frac{\pm \frac{\not\hspace{-0.1mm}P}{2}+\not\hspace{-0.5mm}p+m}
{(\pm \frac{P}{2}+p)^2-m^2+i\epsilon}
\cong 
\frac{1 \pm \gamma^0}{E^B \pm 2p_0-\frac{1}{m}{\vec p}^2+i\epsilon}, 
\label{large1}
\end{eqnarray}
%%%%%%%%%%%%%%%%%%%%%%%%%%%%%%%%%%%%%%%%%%%%%%%%%%%%%%%%%%%%%%%%%%%%%%%%%%%%%
in the large limit of the fermion mass.  
${\vec p}^2/m^2$ term must be kept in the equation since this term may play an 
important role for the convergence of the integration over ${\vec p}$ in the 
right-hand side of Eq. (\ref{BS_Gamma}).  
The amputated Bethe-Salpeter amplitude $\Gamma$ in the rest frame of bound states can 
be written as 
%%%%%%%%%%%%%%%%%%%%%%%%%%%%%%%%%%%%%%%%%%%%%%%%%%%%%%%%%%%%%%%%%%%%%%%%%%%%%
\begin{eqnarray}
\Gamma(p, P)&=&M(p, P)+N(p, P)\not\hspace{-0.5mm}p+F(p, P)(2m+E^B)\gamma^0 
\nonumber \\
&&+G(p, P)\varepsilon^{i0j}p_i(2m+E^B)\gamma_j \ ,
\label{eq:Gamma_rest}
\end{eqnarray}
%%%%%%%%%%%%%%%%%%%%%%%%%%%%%%%%%%%%%%%%%%%%%%%%%%%%%%%%%%%%%%%%%%%%%%%%%%%%%
by substituting Eq. (\ref{mom}) into Eq. (\ref{Gamma}).  
Using Eqs. (\ref{large1}) and (\ref{eq:Gamma_rest}), we find the non-relativistic 
Bethe-Salpeter equation for the amputated Bethe-Salpeter amplitude $\Gamma$ as 
%%%%%%%%%%%%%%%%%%%%%%%%%%%%%%%%%%%%%%%%%%%%%%%%%%%%%%%%%%%%%%%%%%%%%%%%%%%%%
\begin{eqnarray}
\Gamma(q,P)&=&2e^2\int\frac{d^3p}{(2\pi)^3} 
iD_{\mu\nu}(q-p) \gamma^\mu(1+\gamma^0)\gamma^j\gamma^\nu K_j(p, P) , 
\label{BS_trace}
\end{eqnarray}
%%%%%%%%%%%%%%%%%%%%%%%%%%%%%%%%%%%%%%%%%%%%%%%%%%%%%%%%%%%%%%%%%%%%%%%%%%%%%
where 
%%%%%%%%%%%%%%%%%%%%%%%%%%%%%%%%%%%%%%%%%%%%%%%%%%%%%%%%%%%%%%%%%%%%%%%%%%%%%
\begin{eqnarray}
K_j(p, P)&\equiv& \{ N(p, P)p_j+(2m+E^B)G(p, P)\varepsilon_{i0j}p^i \} 
\frac{1}{K(p, P)},
\label{K_j} \\
K(p, P) &\equiv& (E^B-\frac{1}{m}{\vec p}^2+ i\epsilon)^2 -4p_0^2 \ .
\label{eq:function_Kp}
\end{eqnarray}
%%%%%%%%%%%%%%%%%%%%%%%%%%%%%%%%%%%%%%%%%%%%%%%%%%%%%%%%%%%%%%%%%%%%%%%%%%%%%
Now, we can derive the four coupled integral equations for four unknown functions 
$M, N, F$ and $G$ from Eq. (\ref{BS_trace}).  

After tedious calculations\footnote{The details will be appeared elsewhere.~\cite{MN}}, 
we find that the essential coupled equations are those of $N$ and $G$.  
$M$ and $F$ are calculated from $N$ and $G$.  
Furthermore, we find that the independent unknown function is only $T(q,P)$ defined by 
%%%%%%%%%%%%%%%%%%%%%%%%%%%%%%%%%%%%%%%%%%%%%%%%%%%%%%%%%%%%%%%%%%%%%%%%%%%%%
\begin{eqnarray}
T(q,P) \equiv N(q,P)-i(2m+E^B)G(q,P) \ , 
\label{T}
\end{eqnarray}
%%%%%%%%%%%%%%%%%%%%%%%%%%%%%%%%%%%%%%%%%%%%%%%%%%%%%%%%%%%%%%%%%%%%%%%%%%%%%
which is required to satisfy 
%%%%%%%%%%%%%%%%%%%%%%%%%%%%%%%%%%%%%%%%%%%%%%%%%%%%%%%%%%%%%%%%%%%%%%%%%%%%%
\begin{eqnarray}
T(q,P)=\int\frac{d^3p}{(2\pi)^3} R(q-p) 
(-i{\vec q}\cdot{\vec p}-{\vec q}*{\vec p}) \frac{T(p,P)}{K(p, P)},
\label{BS_T}
\end{eqnarray}
%%%%%%%%%%%%%%%%%%%%%%%%%%%%%%%%%%%%%%%%%%%%%%%%%%%%%%%%%%%%%%%%%%%%%%%%%%%%%
where
%%%%%%%%%%%%%%%%%%%%%%%%%%%%%%%%%%%%%%%%%%%%%%%%%%%%%%%%%%%%%%%%%%%%%%%%%%%%%
\begin{eqnarray}
R(q-p)=4\frac{e^2}{{\vec q}^2} \left\{ \frac{1}{(q-p)^2-\mu^2}
\frac{({\vec q}-{\vec p})^2}{(q-p)^2} 
-\alpha\frac{(q^0-p^0)^2}{(q-p)^4}\right\} \ . 
\label{eq:function_R} 
\end{eqnarray}
%%%%%%%%%%%%%%%%%%%%%%%%%%%%%%%%%%%%%%%%%%%%%%%%%%%%%%%%%%%%%%%%%%%%%%%%%%%%%
The symbol ``$*$'' was defined as 
${\vec q}*{\vec p} \equiv \epsilon^{0ij} q_i p_j$.  
Thus by taking the non-relativistic limit, the number of the independent 
equations reduces from four to one.  
Next we derive the Schr\"{o}dinger equation from the non-relativistic 
Bethe-Salpeter equation (\ref{BS_T}).  
Using the residue theorem, we perform the integration over $p^0$ in 
Eq. (\ref{BS_T}).  
We consider poles in the function $K(p, P)$ defined by Eq. (\ref{eq:function_Kp}) 
and get
%%%%%%%%%%%%%%%%%%%%%%%%%%%%%%%%%%%%%%%%%%%%%%%%%%%%%%%%%%%%%%%%%%%%%%%%%%%%%
\begin{eqnarray}
T(q_0,{\vec q},P)=\frac{1}{4}\int\frac{d^2p}{(2\pi)^2} 
R(q_0-p_0^B,{\vec q}-{\vec p}) 
(-{\vec q}\cdot{\vec p}+i{\vec q}*{\vec p})\frac{T(p_0^B,{\vec p},P)} 
{E^B-\frac{{\vec p}^2}{m}+i\epsilon},
\label{BS_T-Sc}
\end{eqnarray}
%%%%%%%%%%%%%%%%%%%%%%%%%%%%%%%%%%%%%%%%%%%%%%%%%%%%%%%%%%%%%%%%%%%%%%%%%%%%%
where $p_0^B\equiv (E^B-{\vec p}^2/m)/2$.  
If we fix $q_0$ to $q_0^B=(E^B-{\vec q^2/m})/2$ in the both sides of 
Eq. (\ref{BS_T-Sc}) and also neglect ${\vec q}^2/2m$ and ${\vec p}^2/2m$ 
for the large fermion mass $m$ again, $R({\vec q}-{\vec p})$ reduces to a 
simple form as 
%%%%%%%%%%%%%%%%%%%%%%%%%%%%%%%%%%%%%%%%%%%%%%%%%%%%%%%%%%%%%%%%%%%%%%%%%%%%%
\begin{eqnarray}
R(q_0^B-p_0^B,{\vec q}-{\vec p}) \cong 
4\frac{e^2}{{\vec q}^2}\frac{1}{({\vec q}-{\vec p})^2+\mu^2} \ .  
\label{eq:neglect_R}
\end{eqnarray}
%%%%%%%%%%%%%%%%%%%%%%%%%%%%%%%%%%%%%%%%%%%%%%%%%%%%%%%%%%%%%%%%%%%%%%%%%%%%%
Then, as the self-consistent equation, we obtain 
%%%%%%%%%%%%%%%%%%%%%%%%%%%%%%%%%%%%%%%%%%%%%%%%%%%%%%%%%%%%%%%%%%%%%%%%%%%%%
\begin{eqnarray}
T(q_0^B,{\vec q},P)=-e^2\int\frac{d^2p}{(2\pi)^2}\frac{1}{{\vec q}^2} 
\frac{1}{({\vec q}-{\vec p})^2+\mu^2} 
({\vec q}\cdot{\vec p} -i{\vec q}*{\vec p})\frac{T(p_0^B,{\vec p},P)} 
{E^B-\frac{{\vec p}^2}{m}+i\epsilon} \ .
\end{eqnarray}
%%%%%%%%%%%%%%%%%%%%%%%%%%%%%%%%%%%%%%%%%%%%%%%%%%%%%%%%%%%%%%%%%%%%%%%%%%%%%

Now we have to define the Sch\"{o}dinger wave function $\psi^B(p)$ of the 
bound states from $T(p_0^B,{\vec p},T)$.  
It is natural to assume that there exists an effective theory for the bound 
states.  
The Hamiltonian of the effective theory, from which the Schr\"odinger equation 
is derived by a variational principle, should be hermite since an energy 
eigenvalue has to be real.  
The hermiticity decides that the wave function should be defined as 
%%%%%%%%%%%%%%%%%%%%%%%%%%%%%%%%%%%%%%%%%%%%%%%%%%%%%%%%%%%%%%%%%%%%%%%%%%%%%
\begin{eqnarray}
\psi^B(p) \equiv \frac{|{\vec p}| T(p_0^B,{\vec p},P)}
{(E^B-\frac{{\vec p}^2}{m}+i\epsilon)} \ , 
\label{eq:psi}
\end{eqnarray}
%%%%%%%%%%%%%%%%%%%%%%%%%%%%%%%%%%%%%%%%%%%%%%%%%%%%%%%%%%%%%%%%%%%%%%%%%%%%%
up to a multiplicative constant.\footnote{See Ref.~\cite{MN} for the details}  
Finally we have 
%%%%%%%%%%%%%%%%%%%%%%%%%%%%%%%%%%%%%%%%%%%%%%%%%%%%%%%%%%%%%%%%%%%%%%%%%%%%%
\begin{eqnarray}
(E^B-\frac{{\vec q}^2}{m}+i\epsilon)\psi^B(q)
=-e^2\int\frac{d^2p}{(2\pi)^2} \frac{1}{|{\vec q}||{\vec p}|}
\frac{{\vec q}\cdot{\vec p}}{({\vec q}-{\vec p})^2+\mu^2}
\psi^B(p) \ . 
\label{Schredinger_eq}
\end{eqnarray}
%%%%%%%%%%%%%%%%%%%%%%%%%%%%%%%%%%%%%%%%%%%%%%%%%%%%%%%%%%%%%%%%%%%%%%%%%%%%%
Notice that the term with $*$-product drops in Eq. (\ref{Schredinger_eq}) because the wave function depends 
on only $|{\vec p}|$.  
This is the Schr\"{o}dinger equation in the momentum space which we solve in 
Sec. 5 to obtain the binding energy .  

\section{Numerical Analysis}

\subsection{Control parameter}

In the Maxwell-Chern-Simons QED$_3$, the square of the coupling constant $e^2$ 
has the dimension of mass.  
The theory is under a control of dimensionless parameters defined by 
%%%%%%%%%%%%%%%%%%%%%%%%%%%%%%%%%%%%%%%%%%%%%%%%%%%%%%%%%%%%%%%%%%%%%%%%%%%%%
\begin{eqnarray}
\hat{\mu}=\frac{\mu}{e^2},\quad\hat{m}=\frac{m}{e^2}.
\end{eqnarray}
%%%%%%%%%%%%%%%%%%%%%%%%%%%%%%%%%%%%%%%%%%%%%%%%%%%%%%%%%%%%%%%%%%%%%%%%%%%%%
These two parameters, not three, are independent in the theory.  
Further more, by defining the dimensionless momenta $\hat{q}$ and $\hat{p}$, and 
dimensionless binding energy ${\hat E}_B$ as 
%%%%%%%%%%%%%%%%%%%%%%%%%%%%%%%%%%%%%%%%%%%%%%%%%%%%%%%%%%%%%%%%%%%%%%%%%%%%%
\begin{eqnarray}
\vec{{\hat q}}=\frac{{\vec q}}{\mu}, \quad
\vec{{\hat p}}=\frac{{\vec p}}{\mu}, \quad
{\hat E}^B=\frac{E^B}{e^2} ,
\end{eqnarray}
%%%%%%%%%%%%%%%%%%%%%%%%%%%%%%%%%%%%%%%%%%%%%%%%%%%%%%%%%%%%%%%%%%%%%%%%%%%%%
we can rewrite the Schr\"{o}dinger equation (\ref{Schredinger_eq}) to 
%%%%%%%%%%%%%%%%%%%%%%%%%%%%%%%%%%%%%%%%%%%%%%%%%%%%%%%%%%%%%%%%%%%%%%%%%%%%%
\begin{eqnarray}
({\hat E}^B-{\vec{\hat q}}^2\frac{\mu^2}{me^2}+i\epsilon)\psi^B(\hat{q})=
-\frac{1}{2}\int\frac{d^2\hat{p}}{(2\pi)^2} \frac{1}{|\vec{\hat q}||\vec{\hat p}|}
\frac{{\vec{\hat q}}\cdot{\vec{\hat p}}}{({\vec{\hat q}}-{\vec{\hat p}})^2+1}
\psi^B(\hat{p})  .
\label{dimless_Schre2}
\end{eqnarray}
%%%%%%%%%%%%%%%%%%%%%%%%%%%%%%%%%%%%%%%%%%%%%%%%%%%%%%%%%%%%%%%%%%%%%%%%%%%%%
We can see that only the combination of parameters as
%%%%%%%%%%%%%%%%%%%%%%%%%%%%%%%%%%%%%%%%%%%%%%%%%%%%%%%%%%%%%%%%%%%%%%%%%%%%%
\begin{eqnarray}
\frac{\mu^2}{m e^2}=\frac{{\hat \mu}^2}{{\hat m}} ,
\label{parameter}
\end{eqnarray}
%%%%%%%%%%%%%%%%%%%%%%%%%%%%%%%%%%%%%%%%%%%%%%%%%%%%%%%%%%%%%%%%%%%%%%%%%%%%%
fully controls Eq. (\ref{dimless_Schre2}).  
Namely, if one chooses a certain value of $\mu^2/(me^2)$, the dimensionless 
energy ${\hat E}^B$ and the wave function $\psi^B(\hat{q})$ are decided.   
To obtain the eigenenergy $E^B$ from the dimensionless ${\hat E}^B$, we need 
to fix one more parameter, ${\hat m}$ or ${\hat \mu}$.  
In the followings, we draw the results in pictures by fixing ${\hat m}$ to see 
the topological mass dependence of ${\hat E}^B$ and also by fixing ${\hat \mu}$ 
to see the fermion mass dependence.  

\subsection{Topological mass dependence}

What we are most interested in is to find how the topological mass affects 
a property of the bound system.  
It can be seen in the shift of the energy eigenvalue by varying the 
topological mass.   
Especially, it is very important to clarify whether the effect by the 
topological mass is attractive or repulsive and how strong the effect is.    

In Fig. 1, we show the topological mass dependence of the 
dimensionless binding energy $E^B/e^2$.  
The vertical axis indicates the binding energy and the horizontal axis is the 
topological mass $\hat{\mu}$ in the logarithmic scale.  
As the topological mass $\hat{\mu}$ increases, the absolute value of 
the binding energy becomes smaller and the number of the bound states 
decreases.  
We find that there exists a critical value of the topological mass 
$\hat{\mu}$ where the bound states disappear, for each a quantum number.  
For example, the critical value of topological mass $\hat{\mu}$ is numerically  
$44.48-44.49$ in the case of the ground state.  

%%%%%%%%%%%%%%%%% Fig. 1 %%%%%%%%%%%%%%%%%
\begin{figure}[ht]
\epsfysize=8cm
\centerline{\epsfbox{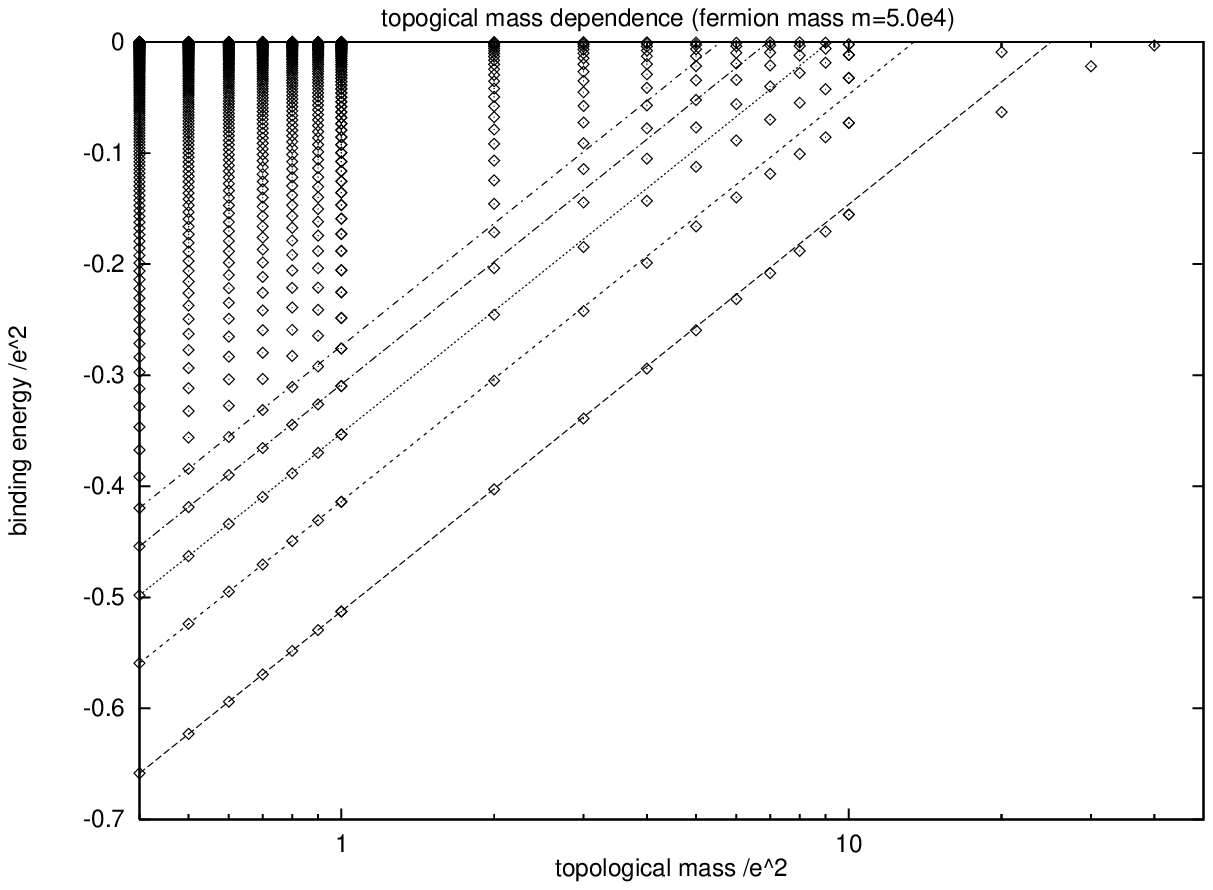}}
%\caption{
Fig. 1. The dependence of the dimensionless binding eigenenergy $E^B/e^2$ 
on the dimensionless topological mass ${\hat \mu}$ with the dimensionless 
fermion mass $\hat{m}=5.0 \times 10^4$.  
Open rhombuses indicate the calculated binding energies $E^B/e^2$.  
The dotted lines are straight with the slope $1/(2\pi)$ as given by 
Eq. (\ref{eq:E^B_log_mu}).
%}
\label{energy_tm}
\end{figure}
%%%%%%%%%%%%%%%%%%%%%%%%%%%%%%%%%%%%%%%%%%%%%%%%%%%%%%%%%%%%

On the other hand, we can see another interesting feature of the topological 
mass dependence of the binding energy from this result.  
In the region where the binding energy is lower than about $-0.3$, the 
topological mass dependence of the binding energy is almost linear in the 
semi-logarithmic figure and the value of its slope is $1/(2\pi)$ at least 
for lower energy states.  
Namely, the dependence of the binding energy seems to be approximated by a 
function form as 
%%%%%%%%%%%%%%%%%%%%%%%%%%%%%%%%%%%%%%%%%%%%%%%%%%%%%%%%%%%%%%%%%%%%%%%%%%%%%
\begin{eqnarray}
\frac{E^B_n}{e^2} = \frac{1}{2\pi} \ln \left(\frac{\mu}{e^2}\right) 
+ C^{(1)}_n,
\label{eq:E^B_log_mu}
\end{eqnarray}
%%%%%%%%%%%%%%%%%%%%%%%%%%%%%%%%%%%%%%%%%%%%%%%%%%%%%%%%%%%%%%%%%%%%%%%%%%%%%
where the index $n$ is a quantum number indicating the energy level.  
$C^{(1)}_n$ is determined so as to reproduce the calculated value of $E^B/e^2$ 
at $\hat{\mu}=0.4$ for each $n$ in Fig. 1.  
Before we discuss about the meaning of this dependence in 
Eq. (\ref{eq:E^B_log_mu}), we show the fermion mass dependence of the binding energy.  
After that, we shall return to this point.  

\subsection{Fermion mass dependence}

In Fig. 2, we show the fermion mass dependence of the binding energy.  
The vertical axis indicates the dimensionless binding energy $E^B/e^2$ and the 
horizontal axis shows the fermion mass $\hat{m}$ in the logarithmic scale.  
The topological mass $\hat{\mu}$ is fixed to be $1.0$.  
As we can see from this figure, the absolute value of the binding energy 
increases as the fermion mass $m$ increases. 
We find the interesting feature again from this figure that the results are 
well approximated by the function as we have seen in the case of the 
topological mass dependence.  
In the region of the binding energy lower than about $-0.3$, the fermion mass 
dependence is almost linear with the slope $1/(4\pi)$ in the semi-logarithmic 
figure.  
Therefore, the fermion mass dependence of the binding energy seems to be 
approximated by the function form as 
%%%%%%%%%%%%%%%%%%%%%%%%%%%%%%%%%%%%%%%%%%%%%%%%%%%%%%%%%%%%%%%%%%%%%%%%%%%%%
\begin{eqnarray}
\frac{E^B_n}{e^2} = -\frac{1}{4\pi}\ln \left(\frac{m}{e^2}\right) 
+ C^{(2)}_n,
\label{eq:E^B_log_m}
\end{eqnarray}
%%%%%%%%%%%%%%%%%%%%%%%%%%%%%%%%%%%%%%%%%%%%%%%%%%%%%%%%%%%%%%%%%%%%%%%%%%%%%
where $C^{(2)}_n$ is determined by the value of $E^B/e^2$ at $m/e^2=5.0 \times 
10^5$ for each $n$ in Fig. 2.  

%%%%%%%%%%%%%%%%% Fig. 2 %%%%%%%%%%%%%%%%%
\begin{figure}[ht]
\epsfysize=8cm
\centerline{\epsfbox{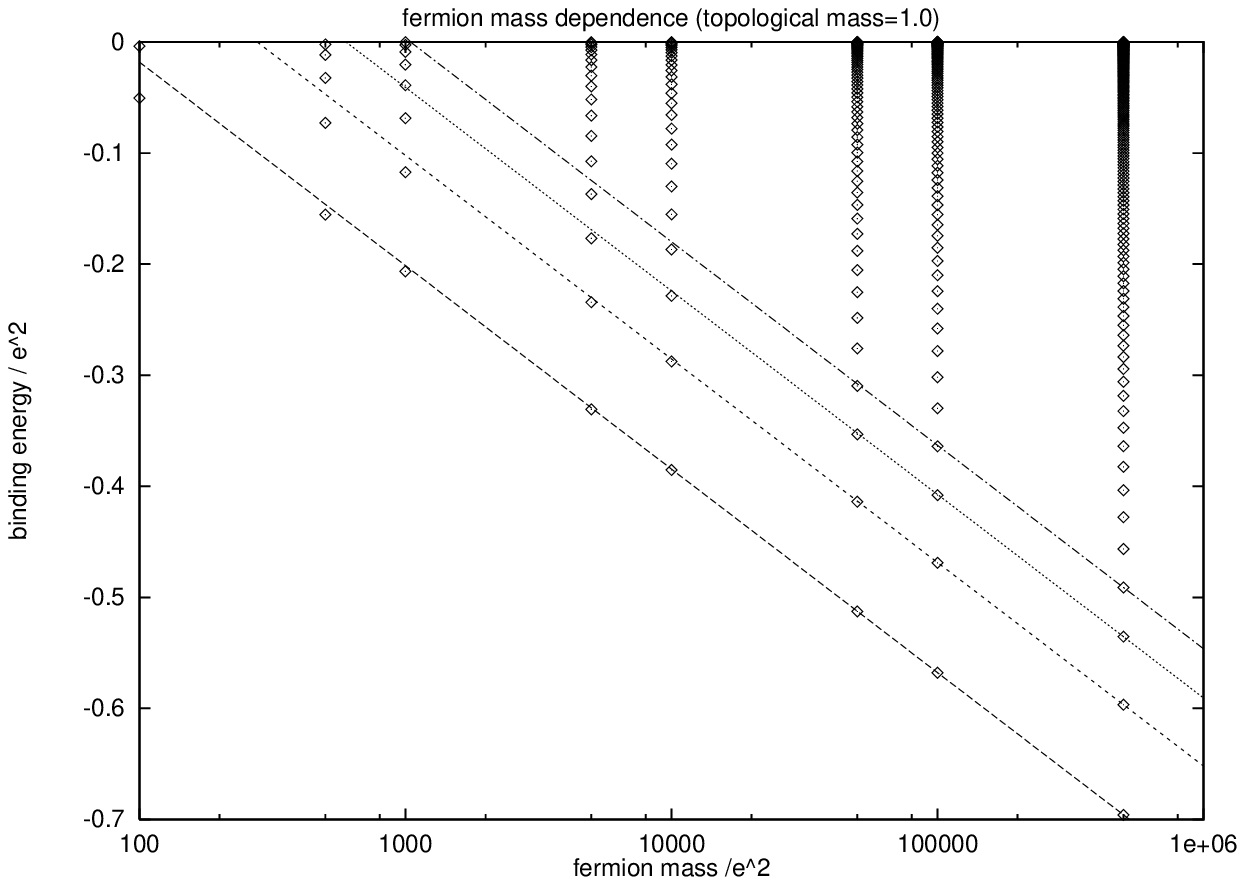}}
%\caption{
Fig. 2. The dependence of the dimensionless binding energy $E^B/e^2$ 
on the dimensionless fermion mass $\hat{m}$ with the dimensionless 
topological mass $\hat{\mu}=1.0$.  
Open rhombuses indicate the calculated binding energies $E^B/e^2$.  
The results can be fitted by the dotted lines defined in 
Eq. (\ref{eq:E^B_log_m}).
%}
\label{energy_fm}
\end{figure}
%%%%%%%%%%%%%%%%%%%%%%%%%%%%%%%%%%%%%%%%%%%%%%%%%%%%%%%%%%%%

\subsection{Logarithmic scaling of binding energy}

Here, we return to the topological mass dependence of the binding energy 
expressed by Eq. (\ref{eq:E^B_log_mu}).  
By combining Eqs. (\ref{eq:E^B_log_mu}) and (\ref{eq:E^B_log_m}), we find that 
the dependence of the binding energy on $\hat{m}$ and $\hat{\mu}$ 
can be unified to the expression as  
%%%%%%%%%%%%%%%%%%%%%%%%%%%%%%%%%%%%%%%%%%%%%%%%%%%%%%%%%%%%%%%%%%%%%%%%%%%%%
\begin{eqnarray}
\frac{E^B_n}{e^2} = \frac{1}{4\pi}\ln \left(\frac{\mu^2}{me^2}\right) 
+ C^{(3)}_n,
\label{eq:E^B_log_mu_m_e}
\end{eqnarray}
%%%%%%%%%%%%%%%%%%%%%%%%%%%%%%%%%%%%%%%%%%%%%%%%%%%%%%%%%%%%%%%%%%%%%%%%%%%%%
where $C^{(3)}_n$ is a constant to be consistent with 
Eqs. (\ref{eq:E^B_log_mu}) and (\ref{eq:E^B_log_m}).  
We can see that the dimensionless binding energy $E^B/e^2$ is completely 
determined by the value of the parameter $\mu^2/(me^2)$ which is 
only one essential parameter included in the theory as seen in Sec. 5.1.

%%%%%%%%%%%%%%%%% Fig. 3 %%%%%%%%%%%%%%%%%
\begin{figure}[ht]
\epsfysize=8cm
\centerline{\epsfbox{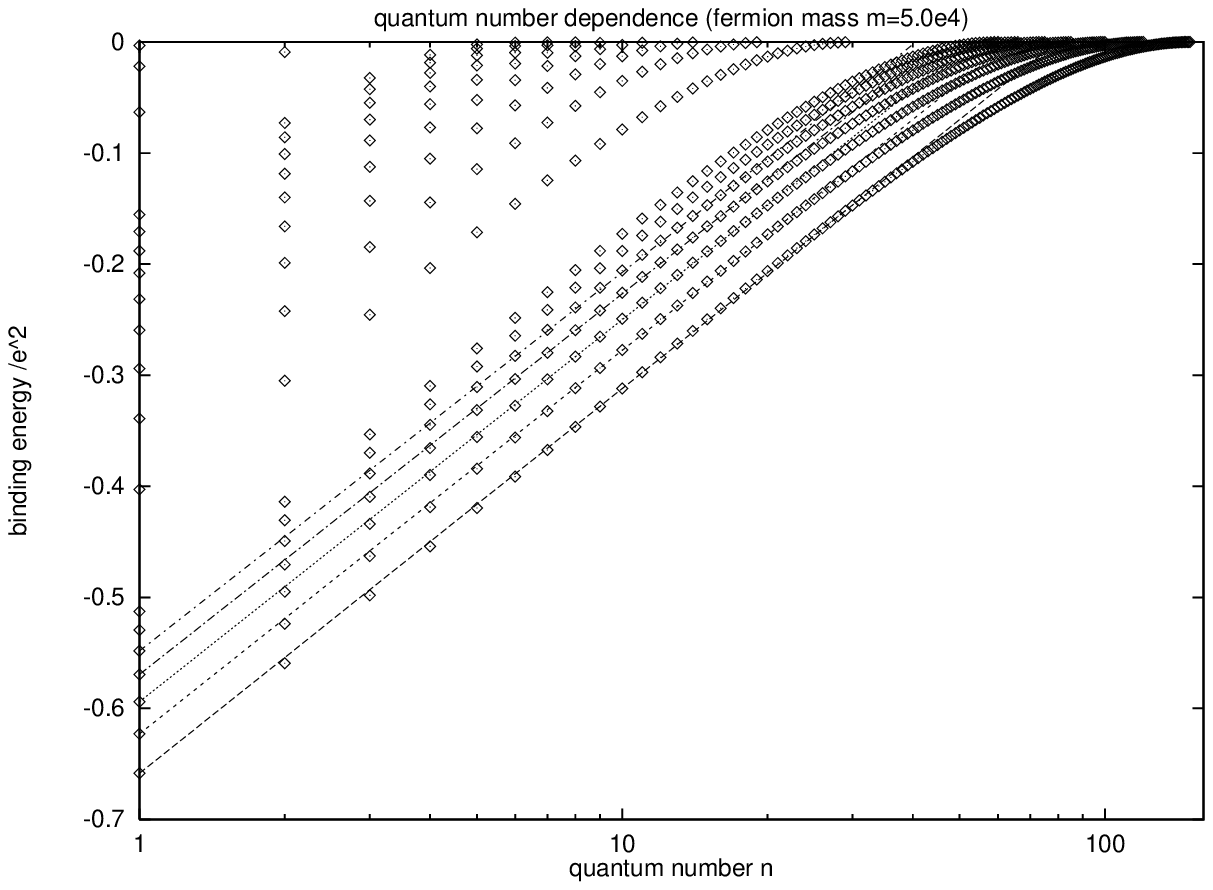}}
%\caption{
Fig. 3. The dependence of the dimensionless binding energy $E^B/e^2$ 
on the quantum number $n$.  
The results can be fitted by the dotted lines defined in 
Eq. (\ref{scaling}) with $a \cong 2.1157 \sim 2.1505$. 
%}
\label{energy_n}
\end{figure}
%%%%%%%%%%%%%%%%%%%%%%%%%%%%%%%%%%%%%%%%%%%%%%%%%%%%%%%%%%%%

In addition, we see the dependence of the binding energy on the quantum 
number $n$ in Fig. 3.  
We can fit the data points for lower energies by a logarithmic function.  
Combining the fitting with Eq. (\ref{eq:E^B_log_mu_m_e}), we obtain 
%%%%%%%%%%%%%%%%%%%%%%%%%%%%%%%%%%%%%%%%%%%%%%%%%%%%%%%%%%%%%%%%%%%%%%%%%%%%%
\begin{eqnarray}
\frac{E^B_n}{e^2} &=& \frac{1}{4\pi}\ln \left(\frac{\mu^2}{me^2}\right)
                   + \frac{1}{a\pi}\ln (n) 
                   =\frac{1}{4\pi}\ln\left(\frac{\mu^2}{me^2} n^{\frac{4}{a}}\right)
\label{scaling}
\end{eqnarray}
%%%%%%%%%%%%%%%%%%%%%%%%%%%%%%%%%%%%%%%%%%%%%%%%%%%%%%%%%%%%%%%%%%%%%%%%%%%%%
where $a$ is a fitting parameter used in Fig. 3.  
Compared with the fittings in Figs. 1 and 2,  the fitting in Fig. 3 results 
us a deviation of the values of $a$ for each fitting line.  
For $n=1 \sim 5$, the fitting parameter varies as $a \cong 2.1157 
\sim 2.1505$.  

The most interesting point of Eq. (\ref{scaling}) is that $E^B_n/e^2$ is 
scaled by the logarithmic function of $\mu^2/(me^2)$ and also $n$.  
If we make a shift as $\mu^2/(me^2) \rightarrow \lambda \mu^2/(me^2)$ or 
$n \rightarrow \lambda n$ where $\lambda$ is a constant, then $E^B_n/e^2$ is 
changed as $E^B_n/e^2 \rightarrow E^B_n/e^2 +1/(4 \pi) \ln \lambda$ or 
$E^B_n/e^2 \rightarrow E^B_n/e^2 +1/(a \pi) \ln \lambda$.  
Thus the logarithmic scaling behaviour means that the multiplicative shift 
of the parameter or the  quantum number results the additive shift 
of the binding energy.  

The scaling has been found only through the numerical analysis in this paper.  
We may expect any simple explanation about the approximated logarithmic 
dependence of $E^B$ on the unique parameter $\mu^2/(me^2)$ and also the energy 
level $n$.   
The coefficient $1/(4\pi)$ is also suggestive for us since it might be a 
signal of a topological nature in the theory.  
These will be studied in future investigations.  

\section{Conclusion}

We have investigated the bound states of the fermion-anti-fermion system in 
the three-dimensional QED with the Chern-Simons term.  
We used the two-component fermion field which belongs to the irreducible 
spinor representation in (2+1)-dimensions.  
We have formulated the Bethe-Salpeter equation which makes it possible to 
perform the systematic  non-perturbative evaluation based on the first 
principle.  
We have performed the calculations in the non-relativistic limit constructing 
the Schr\"{o}dinger equation in the momentum space under the large fermion 
mass expansion.  
By solving this Schr\"{o}dinger equation numerically, we investigated how the 
Chern-Simons term affects to the fermion-anti-fermion bound states.  

We have discovered some important features of the effects of the Chern-Simons 
term in the numerical analysis.  
First, we have found that the absolute value of the binding energy becomes 
smaller as the topological mass $\mu$ larger.  
This fact indicates that the effect of the Chern-Simons term is to weaken 
the binding of the fermion-anti-fermion bound states.  
We have found that there exists a critical value of the topological mass where 
the bound state disappears.  

Secondly, we have found the interesting features of the binding energy that in 
the region where the energy eigenvalue is smaller than about $-0.3$, 
the dependence of the binding energy $E^B_n$ on the topological mass $\mu$, 
the fermion mass $m$ and the energy level $n$ is almost linear in the 
semi-logarithmic plots.  
We find that this dependence can be expressed as 
$E^B_n=e^2/(4\pi)\log(\mu^2 /(me^2) \times n^{4/a})$ with $a \cong 2.1157 
\sim 2.1505$ for the energy levels $n=1 \sim 5$.  
Thus the binding energy has a logarithmic scaling behaviour for varying the 
topological mass, fermion mass and also the quantum number.   

In future investigations, first we should make clear the meanings of the 
logarithmic scaling, the coefficient $1/(4\pi)$ and the critical value of 
the topological mass.  
We should try to obtain any analytical information which helps us to 
understand the meanings of these behaviors of the numerical results.  
Further, we can investigate this problem using the relativistic formulation 
itself without making the non-relativistic approximation.  
We will develop the combined treatment of the Schwinger-Dyson equation 
and the Bethe-Salpeter equation in order to incorporate the effect of the 
dynamical mass generation.  
In addition, we think that one of the most interesting subjects which should 
be done in future is to extend this study to the bound states of the 
fermion-fermion system~\cite{K,B}.  
This is important due to the possible connection to the high-$T_C$ 
superconductivity observed in the condensed matter systems.  

\section*{Acknowledgment}
One of the authors (T. M.) would like to thank Professor M. Kenmoku for his 
hospitality at the Nara Women's University.  
We would like to thank Professor S. Hirenzaki for useful comments and 
advises.  
We also thank to Professor I.I. Kogan for an information on earlier related 
works, and to Professor O.M. Del Cima for an encouraging comment on a future 
development of our analyses.


\begin{thebibliography}{999}
\bibitem{SS}
         W. Siegel, Nucl. Phys. {\bf B156}, 135(1979);
         J. Schonfeld, $ibid$. {\bf B185}, 157(1981).
\bibitem{DJT} 
         S. Deser, R. Jackiw, and S. Templeton, Ann. Phys. (N.Y.) {\bf 140}, 
         372(1982).
\bibitem{RNSI} 
         A. N. Redlich, Phys. Rev. {\bf D29}, 2366(1984); Phys. Rev. Lett. 
         {\bf 52}, 18(1984); 
         A. J. Niemi and G. W. Semenoff, Phys. Rev. Lett. {\bf 51}, 
         2077(1983); 
         K. Ishikawa, $ibid$. {\bf 53}, 1615(1984).
\bibitem{TJZW} See for example: S. B. Treiman, R. Jackiw, B. Zumino and 
         E. Witten, 
         {\it Current Algebra and Anomalies}, (World Scientific, Singapore, 
         1995).
\bibitem{QH} K. von Klitzing, G. Dorda, and M. Pepper. Phys. Rev. Lett. 
             {\bf 45}, 494(1980); 
             D. C. Tsui, H. L. St\"ormer, and A. C. Gossard, Phys. Rev. 
             Lett. {\bf 48}, 1559(1982).  
\bibitem{HTC} G. Bednorz and K. A. M\"uller, Z. Phys. {\bf 64}, 188(1986). 
\bibitem{IM} K. Ishikawa and T. Matsuyama, Z. Phys. {\bf C33}, 41(1986); 
             Nucl. Phys. {\bf B280}[FS18], 523(1987);  
             T. Matsuyama, Prog. Theor. Phys. {\bf 77}, 711(1987); 
             N. Imai, K. Ishikawa, T. Matsuyama, and I. Tanaka, Phys. Rev. 
             {\bf B42}, 10610(1990).
\bibitem{WSM} F. Wilczek, Phys. Rev. Lett. {\bf 48}, 1144(1982);{\bf 49}, 
              957(1982); G.W. Semenoff, Phys. Rev. Lett. {\bf 61}, 517(1988); 
              T. Matsuyama, Phys. Lett. {\bf 228B}, 99(1989); Phys. Lett. 
              {\bf 144A}, 59(1990); Jour. Phys. {\bf A23}, 5241(1990); 
              Phys. Rev. {\bf D42}, 3469(1990); Prog. Theor. Phys. {\bf 84}, 
              1220(1990).
\bibitem{JTAPRS} 
         R. Jackiw and S. Templeton, Phys. Rev. {\bf D23}, 2291(1981); 
         T. Appelquist and R. D. Pisarski, $ibid$. {\bf D23}, 2305(1981); 
         S. Templeton, $ibid$. {\bf D24}, 3134(1981);
         M. de Roo and K. Stam, Nucl. Phys. {\bf B246}, 335(1984).
\bibitem{PABCWABKW} 
         R. D. Pisarski, Phys. Rev. D {\bf 29}, 2423(1984); 
         T. Appelquist, M. J. Bowick, E. Cohler, and L. C. R. Wijewardhana, 
         Phys. Rev. Lett. {\bf 55}, 1715(1985); 
         T. Appelquist, M. J. Bowick, D. Karabali and L. C. R. Wijewardhana, 
         Phys. Rev. {\bf D33}, 3704(1986).
\bibitem{DKK}
         E. Dagotto, J.B. Kogut and A. Koci\'c, Phys. Rev. Lett. {\bf 62}, 
         1083(1989).
\bibitem{KKKM}
         W.-H. Kye and J. K. Kim, Phys. Rev. {\bf D50}, 5398(1994); 
         K.-I. Kondo and P. Maris, Phys. Rev. Lett. {\bf 74}, 18(1995); 
         $ibid.$ Phys. Rev. {\bf D52}, 1212(1995); 
         K.-I. Kondo, T. Ebihara, T. Iizuka and E. Tanaka, Nucl. Phys. 
         {\bf B434}, 85(1995).
\bibitem{HMH}
         Y. Hoshino and T. Matsuyama, Phys. Lett. B{\bf 222}, 493(1989); 
         see also 
         Y. Hoshino, T. Matsuyama and C. Yoshida-Habe, in {\it Proceedings of 
         the 1989 Workshop on Dynamical Symmetry Breaking, Nagoya, Japan}, 
         edited by T. Muta and K. Yamawaki (Nagoya University, Nagoya, 1990). 
\bibitem{MNU}
          T. Matsuyama, H. Nagahiro and S. Uchida, Phys. Rev. {\bf D60}, 
          105020(1999); 
          T. Matsuyama and H. Nagahiro, Gravitation and Cosmology {\bf 6}, 
          145(2000); {\it ibid.} Mod. Phys. Lett. {\bf A15}, 2373(2000).  
\bibitem{BAB}
          C.J. Burden, Nucl. Phys. {\bf B387}, 419(1992); 		  
    	  T. W. Allen and C.J. Burden, Phys. Rev. {\bf D53}, 5842(1996); 
	  {\it ibid.} (E) {\bf D54}, 6567(1996); {\it ibid.} {\bf D55}, 
	  4954(1997).  
\bibitem{K} Ya.I. Kogan, JETP Lett. {\bf 49}, 225(1989); see also I.I. Kogan, 
            Phys. Lett. {\bf B262}, 83(1991); I.I. Kogan and G.W. Semenoff, 
            Nucl. Phys. {\bf B368}, 718(1992); R.J. Szabo, I.I. Kogan and 
            G.W. Semenoff, {\it ibid} {\bf B392}, 700(1993). 
\bibitem{GGS} 
          H. O. Girotti, M. Gomes and A. J. da Silva,
          Phys. Lett. B {\bf 274},  357 (1992); 
          H.G. Girotti, M. Gomes, J.L. deLyra, R.S. Mendes, J.R.S. Nascimento, 
	  and A.J. da Silva, Phys. Rev. Lett. {\bf 69}, 2623(1992);{\it ibid.}
	  {\bf 71}, 203(1993).   
\bibitem{H} C. R. Hagen, Phys. Rev. Lett. {\bf 71}, 202(1992). 
\bibitem{DEKSS}          
      M.I. Dobroliubov, D. Eliezer, I.I. Kogan, G.W. Semenoff and R.J. Szabo, 
      Mod. Phys. Lett. {\bf A8}, 2177(1993). 
\bibitem{CS}
          S.S. Chern and J. Simons, Ann. Math. {\bf 99}, 48(1974).
\bibitem{BS}
          E. E. Salpeter, H. A. Bethe, Phys. Rev. {\bf 84}, 1232(1951).
\bibitem{L}
          David Lurie, ``Particles and Fields'', 
          Interscience Publishers (1968).
\bibitem{MN}
          T. Matsuyama and H. Nagahiro, to be appeared.  
\bibitem{B} H. Belich, O.M. Del Cima, M.M. Ferreira Jr. 
            and J.A. Helay\"el-Neto, 
            Int. J. Mod. Phys. {\bf A16}, 4939(2001).
\end{thebibliography}
\end{document}